\documentclass[journal, onecolumn, draftcls]{IEEEtran}
\ifCLASSINFOpdf
\else
\fi
\usepackage[cmex10]{amsmath}
\usepackage{amsfonts}
\usepackage{amssymb}
\usepackage{bm}
\usepackage{dsfont}
\usepackage{xcolor}
\usepackage{yfonts}

\newtheorem{theorem}{Theorem}

\newtheorem{remark}[theorem]{Remark} 

\newcommand{\bra}[1]{\langle #1|}
\newcommand{\ket}[1]{|#1\rangle}
\newcommand{\braket}[2]{\langle #1|#2\rangle}
\DeclareMathOperator{\Tr}{Tr}
\DeclareMathOperator*{\argmin}{arg\,min}

\newcommand{\Esp}{E_{\text{sp}}}
\newcommand{\Espcc}{E_{\text{sp}}^{\text{cc}}}

\newcommand{\Pe}{\mathsf{P}_{\text{e}}}
\newcommand{\Pem}{\mathsf{P}_{\text{e}|m}}
\newcommand{\Pemax}{\mathsf{P}_{\text{e,max}}}
\newcommand{\channel}{\textswab{C}}

\newcommand{\Code}{\bm{\mathcal{C}}}

\newcommand{\dmin}{d_{\text{min}}}

\begin{document}
%
\title{Constant Compositions in the  Sphere Packing Bound for Classical-Quantum Channels}


\author{Marco~Dalai,~\IEEEmembership{Member,~IEEE}, Andreas~Winter
\thanks{M. Dalai is with the Department
of Information Engineering, University of Brescia, Italy, email: marco.dalai@ing.unibs.it.

A. Winter is with ICREA \& F\'isica Te\`orica: Informaci\'o i Fenomens Qu\`antics, Universitat Aut\`{o}noma de Barcelona, Spain,
email: andreas.winter@uab.cat

Part of the results where first presented in \cite{dalai-winter-2014}.
}
}


%


\maketitle

\begin{abstract}

The sphere packing bound, in the form given by Shannon, Gallager and Berlekamp, was recently extended to classical-quantum channels, and it was shown that this creates a natural setting for combining probabilistic approaches with some combinatorial ones such as the Lov\'asz theta function.
In this paper, we extend the study to the case of constant composition codes. We first extend the sphere packing bound for classical-quantum channels to this case, and we then show that the obtained result is related to a variation of the Lov\'asz theta function studied by Marton. We then propose a further extension to the case of varying channels and codewords with a constant  conditional composition given a particular sequence. This extension is then applied to auxiliary channels to deduce a bound which can be interpreted as an extension of the Elias bound.
\end{abstract}


%
\IEEEpeerreviewmaketitle

\section{Introduction}








The sphere packing  bound has been recently extended to classical-quantum channels \cite{dalai-ISIT-2012}, \cite[Sec. V]{dalai-TIT-2013} by resorting to the first rigorous proof given for the case of classical discrete memoryless channels (DMC) by Shannon, Gallager and Berlekamp \cite{shannon-gallager-berlekamp-1967-1}.
That resulted in an upper bound to the reliability function of classical-quantum channels, which is the error exponent achievable by means of optimal codes.

The classical proof given in \cite{shannon-gallager-berlekamp-1967-1} can be considered a rigorous completion of Fano's first efforts toward proving the bound \cite[Ch. 9]{fano-book}. However, while Fano's approach led to a tight exponent at high rates for general constant composition codes, the proof in \cite{shannon-gallager-berlekamp-1967-1} only considers the case of the optimal composition.
Shortly afterwards, Haroutunian \cite{haroutunian-1968}, \cite{csiszar-korner-book}, proposed a simple yet rigorous proof which gives the tight exponent for codes with general (possibly non optimal) constant composition. However, a greedy extension of this proof to classical-quantum channels does not give a good bound (see \cite[Th. II.20 and page 35]{winter-phd-1999}). This motivated the choice made in \cite{dalai-ISIT-2012, dalai-TIT-2013} to follow the approach of \cite{shannon-gallager-berlekamp-1967-1}.

In this paper, we modify slightly the approach in \cite{dalai-ISIT-2012, dalai-TIT-2013}  to derive a sphere packing bound for classical-quantum channels with constant composition codes. The main difference with respect to the classical case is in the resulting possible analytical expressions of the bound, which does not seem to be expressible, in this case, in terms of the Kullback-Leibler divegence and mutual information.
In analogy with the results obtained in \cite{dalai-ISIT-2013a} \cite[Sec. VI]{dalai-TIT-2013}, we then discuss the connections of the constant composition version of the bound with a quantity  introduced by Marton \cite{marton-1993} as a generalization of the Lov\'asz theta function for bounding the highest rate achievable by zero-error codes with codewords of a given arbitrary composition.
Finally, we propose an extension of the sphere packing bound for varying channels and codewords with a constant \emph{conditional} composition from a given sequence, and we show that this result includes as a special case a recently developed generalization of the Elias bound \cite{dalai-ISIT-2014a}.

\section{Definitions}
Consider a classical-quantum channel $\channel$ with input alphabet $ \mathcal{X}=\{1,\ldots,| \mathcal{X}|\}$ and associated density operators $S_ x$, $ x\in \mathcal{X}$, in a finite dimensional Hilbert space $\mathcal{H}$. The $n$-fold product channel acts in the tensor product space $\bm{\mathcal{H}}=\mathcal{H}^{\otimes n}$ of $n$ copies of $\mathcal{H}$. To a sequence $\bm{ x}=( x_1, x_2,\ldots, x_n)$ we associate the signal state $\bm{S}_{\bm{ x}}=S_{ x_1}\otimes S_{ x_2}\cdots\otimes S_{ x_n}$.
A block code with $M$ codewords is a mapping from a set of $M$ messages $\{1,\ldots,M\}$ into a set of $M$ codewords  $\bm{ x}_1,\ldots, \bm{ x}_M$ and the rate of the code is $R=(\log M)/n$.

We consider a quantum decision scheme for such a code (POVM) composed of a  collection of $M$ positive operators $\{\Pi_1,\Pi_2,\ldots,\Pi_M\}$ such that $\sum \Pi_m \leq \mathds{1}$, where $\mathds{1}$ is the identity operator. 
The probability that message $m'$ is decoded when message $m$ is transmitted is $\mathsf{P}_{m'|m}=\Tr \Pi_{m'} \bm{S}_{\bm{ x}_m}$ and the probability of error after sending message $m$ is
\begin{equation*}
\Pem=1-\Tr\left(\Pi_m \bm{S}_{\bm{ x}_m}\right).
\end{equation*}
The maximum error probability of the code is defined as the largest $\Pem$, that is,
\begin{equation*}
\Pemax=\max_{m}\Pem.
\end{equation*}
In this paper, we are interested in bounding the probability of error for constant composition codes. Given a composition $P_n$, we define  $\Pemax^{(n)}(R,P_n)$ to be the smallest maximum error probability among all codes of length $n$, rate \emph{at least} $R$, and composition $P_n$.
For a probability distribution $P$, we define the asymptotic optimal error exponent with composition $P$ as
\begin{equation}
E(R,P)=\limsup_{n\to\infty} -\frac{1}{n}\log \Pemax^{(n)}(R_n,P_n),
\label{eq:E(R)_def_class}
\end{equation} 
where the limsup is over all sequences of codes with rates $R_n$ and compositions $P_n$ such that $R_n \to R$ and $P_n \to P$ as $n\to\infty$. For channels with a zero-error capacity, the function $E(R,P)$ can be infinite for rates $R$ smaller than some given quantity $C_0(P)$, which we can call the zero-error capacity of the channel relative to $P$. It is important to observe that, as for $C_0$, the value $C_0(P)$ only depends on the confusability graph $G$ of the channel, for which we could also call it $C(G,P)$ \cite{csiszar-korner-1981}, \cite{marton-1993}. 

To avoid unnecessary complications, we use a flexible notation in this paper. We keep it simple as far as possible, progressively increasing its complexity by adding arguments to functions as their definitions become more general. The meaning of all quantities will be clear from the context.

\section{Sphere Packing Bound for Constant Composition Codes}
Our main result is the following theorem.
\begin{theorem}
\label{th:cc-sphere-packing}
For all positive rates $R$, distribution $P$, and positive $\varepsilon < R$, we have the bound
\begin{equation*}
E(R,P)\leq \Espcc(R-\varepsilon, P),
\end{equation*}
where $\Espcc(R,P)$ is defined by the relations
\begin{align}
\Espcc(R,P) & =  \sup_{\rho \geq 0} \left[ E_0^{\text{cc}}(\rho,P) - \rho R\right]
,\\
E_0^{\text{cc}}(\rho,P) & = \min_{F} \left[-(1+\rho)\sum_x P(x) \log\Tr(S_x^{\frac{1}{1+\rho}}F^{\frac{\rho}{1+\rho}})\right].
\label{eq:E0ccrhoP}
\end{align}
the minimum being over all density operators $F$.
\end{theorem}

\begin{IEEEproof}
See Appendix \ref{app:proof_sp}.
\end{IEEEproof}

The bound is written here in terms of R\'enyi divergences. For commuting states, that is, classical channels, the bound can be written in the more usual form in terms of Kullback-Leibler  divergences and mutual information as in \cite{csiszar-korner-book}.
In fact, assuming that the states $S_x$ commute, let for notational convenience $W(y|x)$ be their eigenvalues, which we interpret as classical probability distributions, indexing in $y$ the output space. Then we can write (see \cite[Ch. 5, Prob. 23]{csiszar-korner-book})
\begin{align}
E_0^{\text{cc}}(\rho,P) & = \min_{F} \left[-(1+\rho)\sum_x P(x) \log\Tr(S_x^{\frac{1}{1+\rho}}F^{\frac{\rho}{1+\rho}})\right]\\
& = \min_{Q} \left[-(1+\rho)\sum_x P(x) \log\sum_y W(y|x)^{\frac{1}{1+\rho}}Q(y)^{\frac{\rho}{1+\rho}})\right]\\
& = \min_{V,Q}\sum_{x,y}P(x)V(y|x)\left[\log\frac{V(y|x)}{W(y|x) } +\delta \log \frac{V(y|x)}{Q(y)} \right]\\
& = \min_{V} \left[ D(V||W|P)+\delta I(P,V) \right],
\end{align}
where the $V(\cdot|x)$ and $Q$ run over probability distributions on $y$, $I(P,V)$ is the mutual information with the notation of \cite{csiszar-korner-book} 
\begin{equation}
I(P,V)=\sum_{x,y} P(x)V(y|x) \log \frac{V(y|x)}{\sum_{x'}P(x')V(y|x')},
\end{equation}
and $D(V||W|P)$ is the conditional information divergence
\begin{equation}
D(V||W|P)=\sum_{x}P(x)\sum_y V(y|x)\log \frac{V(y|x)}{W(y|x)}.
\end{equation}
Hence, for classical channels, we have the more familiar form of the bound (see \cite{csiszar-korner-book})
\begin{align}
\Espcc(R,P) & =  \sup_{\rho \geq 0} \left[ \min_{V} \left( D(V||W|P)+\delta I(P,V) \right)- \rho R\right]\\
& = \min_{V:I(P,V)\leq R} D(V||W|P).
\label{eq:EspKLdivergenceCL}
\end{align}
This form of the bound emerges naturally in Haroutunian's proof \cite{haroutunian-1968}, \cite{csiszar-korner-book}, which is very simple and gives a very intuitive interpretation of the resulting expression. For a given rate $R$, one considers auxiliary channels $V$ such that $I(P,V)<R$. Given codes with rate $R$ and composition $P$, by the strong converse to the coding theorem, the probability of error over channel $V$ for at least one codeword is nearly one. For that same codeword, the probability of error over channel $W$ can be lower bounded in terms of the Kullback-Leibler divergence $D(V||W|P)$, and this leads to the sphere packing bound.

It is interesting to consider what happens in the case of non-commuting states. A reasoning similar to the one described in the last paragraph can be applied to derive a bound which is the formal analog of the classical one in the form given using equation \eqref{eq:EspKLdivergenceCL}, namely  (see \cite[Th. II.20]{winter-phd-1999})
\begin{equation}
E(R,P)\leq \min_{V:I(P,V)\leq R} D(V||S|P)
\label{eq:haroutunianQ}
\end{equation}
where now the minimum is over all set of density operators $V_x$, 
\begin{equation}
I(P,V)=H\left(\sum_x P(x)V_x\right)-\sum_x P(x)H(V_x), \quad \mbox{with }H(\rho)=-\Tr\rho\log \rho,
\end{equation}
and
\begin{equation}
D(V||S|P)=\sum_x P(x)\Tr V_x(\log V_x-\log S_x).
\end{equation}
The main difference with respect to the classical case, however, is that this bound does not have good properties in the more general classical-quantum setting. For example, note that - as in the classical case - the bound is finite only when the $V_x$ can be chosen so that $\mbox{supp}(V_x)\subseteq \mbox{supp}(S_x)$. As a consequence, for pure-state channels the bound is infinite for rates $R<I(P,S)$, which means that the bound is essentially trivial in this case. The reason for this unexpected behavior can be traced back to a fundamental difference in the study of error exponents in the classical and quantum binary hypothesis testing (see for example \cite[Sec. 4.8]{audenaert-et-al-2008}). A more detailed discussion of this issue requires an inspection of the proof of the sphere packing bound and is thus deferred to Appendix \ref{app:BHT-cl-cl-q}.

Now it is not difficult to show that after optimization of the composition we recover the original bound of \cite{dalai-ISIT-2012}, \cite{dalai-TIT-2013}. In order to do this, note that 
\begin{align*}
\max_P \Espcc(R) & =  \sup_{\rho \geq 0} \left[ \max_P E_0^{\text{cc}}(\rho,P) - \rho R\right].
\end{align*}
Then, 
\begin{align*}
\max_P  & E_0^{\text{cc}}(\rho,P) \\
&  = \max_P \min_{F} \left[-(1+\rho)\sum_x P(x) \log\Tr(S_x^{\frac{1}{1+\rho}}F^{\frac{\rho}{1+\rho}})\right].\\
& =  \min_{F} \max_P  \left[-(1+\rho)\sum_x P(x) \log\Tr(S_x^{\frac{1}{1+\rho}}F^{\frac{\rho}{1+\rho}})\right]\\
& =  \min_{F} \left[-(1+\rho)\max_x \log\Tr(S_x^{\frac{1}{1+\rho}}F^{\frac{\rho}{1+\rho}})\right],
\end{align*}
where the minimum and the maximum can be exchanged due to linearity in $P$ and convexity in $F$. The resulting expression is in fact the coefficient $E_0(\rho)$ which defines the sphere packing bound as proved in \cite[Th.~6]{dalai-TIT-2013}. Hence, this procedure allows us to recover the results of \cite{dalai-ISIT-2012}, \cite{dalai-TIT-2013} by noticing that
\begin{align}
E(R) & = \sup_{P} E(R,P) \\
&\leq \sup_{P} \Espcc(R-\varepsilon,P)\\
&= \Esp(R-\varepsilon).
\end{align}
Theorem \ref{th:cc-sphere-packing} constitutes thus the most general form of the sphere packing bound, from which all other forms can be derived.

\section{Connections with Marton's function}
The bound $\Espcc(R,P)$ obtained in the previous section can be used as an upper bound for the zero-error capacity of the channel relative to $P$. Whenever the function $\Espcc(R-\varepsilon,P)$ is finite, in fact, then the probability of error at rate $R$ is non-zero. It is not difficult to observe that the smallest rate $R_\infty(P)$ at which $\Espcc(R,P)$ is finite can be evaluated as
\begin{align*}
R_\infty(P) & = \lim_{\rho\to\infty} \frac{E_0^{\text{cc}}(\rho,P)}{\rho}\\
& = \min_{F} \left[-\sum_x P(x) \log\Tr(S_x^0 F)\right],
\end{align*}
where $S_x^0$ is the projection onto the range of $S_x$. When optimized over $P$, we obtain the expression
\begin{align*}
R_\infty & = \min_{F} \max_x \log\frac{1}{\Tr(S_x^0 F)},
\end{align*}
already discussed in \cite{dalai-TIT-2013}. Hence, we have the bounds $C_0(P)\leq R_\infty(P)$ and $C_0\leq R_\infty$.

It was observed in \cite{dalai-ISIT-2013a} and \cite[Sec. VI]{dalai-TIT-2013} that $R_\infty$ is related to the Lov\'asz number $\vartheta$ \cite{lovasz-1979}.
Here, we observe that, in complete analogy, the value $R_\infty(P)$ is related to a variation of the $\vartheta$ function introduced by Marton in \cite{marton-1993} as an upper bound to $C(G,P)$.
Given a (confusability) graph $G$, Marton introduces the following quantity\footnote{We use the notation $\vartheta(G,P)$ in place of Marton's $\lambda(G,P)$ to preserve a higher coherence with the context of this paper. For the same reason, in what follows we also use, as in \cite{dalai-TIT-2013},  a logarithmic version of the ordinary Lov\'asz $\vartheta$ function, that is, our $\vartheta$ corresponds to $\log\vartheta$ in Lov\'asz' notation. \label{note:loglovasz}}:
\begin{equation}
\vartheta(G,P)
         = \min_{\{u_x\}, f} \sum_x P(x) \log\frac{1}{|\braket{u_x}{f}|^2},
\label{eq:martonstheta}
\end{equation}
where the minimum is over all representations $\{u_x\}$ of the graph $G$ in the Lov\'asz sense and over all unit norm vectors $f$. She then shows that $C(G,P)\leq \vartheta(G,P)$.

Let us now compare this bound with the best bound on $C(G,P)$ that we can deduce from the sphere packing bound using $R_\infty(P)$. We enforce the notation writing $R_\infty(\{S_x\}, P)$  to point out the dependence of $R_\infty(P)$ on the channel states $S_x$. For a given confusability graph $G$, the best upper bound to $C(G,P)$ is obtained by minimizing $R_\infty(\{S_x\}, P)$ over all possible  channels with confusability graph $G$. We may then define
\begin{align}
\vartheta_{\text{sp}}(G,P)
     & = \inf_{\{S_x\}} R_\infty(\{S_x\}, P)\\
     & = \inf_{\{U_x\}, F} \sum_x P(x) \log\frac{1}{\Tr(U_x F)},
\label{eq:thetasp}
\end{align}
where $\{U_x\}$ now runs over all sets of projectors with confusability graph $G$. Then we have the bound $C(G,P)\leq \vartheta_{\text{sp}}(G,P)$.

The quantity $\vartheta_{\text{sp}}(G,P)$ is the constant composition analog of the formal quantity $\vartheta_{\text{sp}}(G)$ defined in \cite[Sec.~VI]{dalai-TIT-2013}. 
In that case it was observed by Schrijver and by Duan and Winter \cite{DuanWinter} that in fact $\vartheta_{\text{sp}}(G)=\vartheta(G)$ (with our logarithmic definition of $\vartheta$, see footnote \ref{note:loglovasz}). We have the analogous result for constant compositions.
\begin{theorem}
  For any graph $G$ and composition $P$, $\vartheta_{\text{sp}}(G,P)=\vartheta(G,P)$.
  \label{th:thetaspeqtheta}
\end{theorem}
\begin{IEEEproof}
It is obvious that $\vartheta_{\text{sp}}(G,P)\leq\vartheta(G,P)$, since the right hand side of  \eqref{eq:martonstheta} is obtained by restricting the operators in the right hand side of \eqref{eq:thetasp} to have rank one.

We now prove the converse inequality (cf.~\cite{DuanWinter}). 
Let $\{U_x\}$ and $F$ be a representation of $G$ and a state respectively. 
Let first $\ket{\psi}\in \mathcal{H}\otimes \mathcal{H}'$ be a purification of $F$ obtained using an auxiliary space $\mathcal{H}'$, so that $\Tr(U_x F)=\Tr(U_x\otimes \mathds{1}_{\mathcal{H}'}\ket{\psi}\bra{\psi})$. Let then 
\begin{equation}
\ket{w_x}=\frac{U_x\otimes \mathds{1}_{\mathcal{H}'}\ket{\psi}}{\|U_x\otimes \mathds{1}_{\mathcal{H}'}\ket{\psi}\|}.
\end{equation}
It is not difficult to check that $\{w_x\}$ is an orthonormal representation of $G$ and that 
$\Tr(U_x F) = \Tr(U_x\otimes \mathds{1}_{\mathcal{H}'}\ket{\psi}\bra{\psi})
            = |\braket{w_x}{\psi}|^2$, for all $x$.
Hence, the orthormal representation $\{w_x\}$ and the unit norm vector $\psi$ satisfy
\begin{equation}
  \sum_x P(x) \log\frac{1}{\Tr(U_x F)}
         = \sum_x P(x)\log\frac{1}{|\braket{w_x}{\psi}|^2},
\end{equation}
which implies that $\vartheta(G,P) \leq \vartheta_{\text{sp}}(G,P)$.
\end{IEEEproof}

We can now discuss another interesting issue about the use of the quantity $\vartheta(G,P)$. When we are interested in bounding $C_0$, we can use the bound $C_0\leq \vartheta(G)$ or we can also use the bound\footnote{Note that $C_0 = \max_P C_0(P)$, since the number of compositions is polynomial in the block-length.}  $C_0\leq \max_P \vartheta(G,P)$.
Marton \cite{marton-1993} states that this does not make a difference since - ``as is easily seen''  - $\max_P \vartheta(G,P)=\vartheta(G)$. However, a proof of this statement does not seem to follow easily from the definitions. It can in fact be written as
\begin{align}
\max_P \min_{\{u_x\}, f} \sum_x P(x) \log\frac{1}{|\braket{u_x}{f}|^2} & = 
\min_{\{u_x\}, f} \max_x  \log\frac{1}{|\braket{u_x}{f}|^2}\\
& = \min_{\{u_x\}, f} \max_P  \sum_x P(x) \log\frac{1}{|\braket{u_x}{f}|^2}
\end{align}
and, in order to prove the equality, we would need to exchange the maximization over $P$ with the minimization over representations and handles. It is not clear in Marton's paper what argument she used to motivate it.
We use Theorem \ref{th:thetaspeqtheta} to prove this statement.
\begin{theorem}
  For any graph $G$, $\max_P \vartheta(G,P)=\vartheta(G)$.
\end{theorem}
\begin{IEEEproof}
For any representation $\{U_x\}$ of $G$ and density operator $F$, define
the function $f(x) = \Tr U_x F$, and denote the set of all functions $f$
obtained in this way by $\text{OR}(G)$.
The proof of Theorem \ref{th:thetaspeqtheta} shows that any $f\in\text{OR}(G)$
can be realized by rank-one projections $U_x = \ket{u_x}\bra{u_x}$
and a pure state $F = \ket{f}\bra{f}$, in a space of dimension at most
$|\mathcal{X}|$ (namely the span of the $\ket{u_x}$). 
In particular, it follows that $\text{OR}(G)$ is closed and compact.

Furthermore, it is convex: namely, consider $f_i(x) = \Tr U_x^{(i)} F^{(i)}$
for representations $\{U_x^{(i)}\}$ of $G$ and density operators $F^{(i)}$,
$i=1,2$. Then, for $0\leq p \leq 1$, let
$U_x = U_x^{(1)} \oplus U_x^{(2)}$ and 
$F = p F^{(1)} \oplus (1-p) F^{(2)}$, which has associated
$f(x) = \Tr U_x F = p f_1(x) + (1-p) f_2(x)$, i.e.~$p f_1 + (1-p) f_2 \in \text{OR}(G)$.

Now define the quantity
\begin{equation}
  J(f,P)=\sum_x P(x)\log \frac{1}{f(x)},
\end{equation}
for compositions $P$ and functions $f\in\text{OR}(G)$.
The theorem is equivalent to the statement that
\begin{equation}
  \max_P \min_{f\in\text{OR}(G)} J(f,P)= \min_{f\in\text{OR}(G)} \max_P J(f,P),
  \label{eq:minimax}
\end{equation}
since the left hand side equals $\max_P \vartheta(G,P)$ by 
Theorem \ref{th:thetaspeqtheta}, 
and the right hand side equals $\vartheta(G)$ by \cite[Th. 8]{dalai-TIT-2013}.

But \eqref{eq:minimax} is an instance of the minimax theorem. Indeed, both
the domains of $f$ and $P$ are convex and compact, and the functional
$J$ is convex in the former and concave (in fact affine linear) in
the latter.
\end{IEEEproof}

We close this section with a simple yet useful result which we will need in the next section. This is the analogous of \cite[Th. 10]{dalai-TIT-2013} for the constant composition setting.
\begin{theorem}
For any pure-state channel we have the inequality $\Espcc(R_\infty(P),P)\leq R_\infty(P)$.
\label{th:Erinfty}
\end{theorem}
\begin{IEEEproof}
For a pure state channel, since $S_x^{\frac{1}{1+\rho}}=S_x=S_x^0$, we have
\begin{align*}
E_0^{\text{cc}}(\rho,P) & = \min_{F} \left[-(1+\rho)\sum_x P(x) \log\Tr(S_x^{\frac{1}{1+\rho}}F^{\frac{\rho}{1+\rho}})\right]\\
&= \min_{F} \left[-(1+\rho)\sum_x P(x) \log\Tr(S_x F^{\frac{\rho}{1+\rho}})\right]\\
& \leq  \min_{F} \left[-(1+\rho)\sum_x P(x) \log\Tr(S_x^0 F)\right]\\
& = (1+\rho) R_\infty(P),
\end{align*}
from which we easily deduce the statement by definition of $\Espcc(R,P)$.
\end{IEEEproof}

\section{Conditional Compositions}
\subsection{Conditional Sphere Packing Bound}
We now develop an extension of the sphere packing to handle the case of varying channels
with a \emph{conditional composition} constraint on the codewords.
Although this setting  can appear artificial, the bound will prove useful when applied to auxiliary channels in a procedure that can be considered as an evolution of the method used in \cite[Sec. VIII]{dalai-TIT-2013} along the same lines taken in \cite{dalai-ISIT-2014a}.
Here we assume that we have a finite set $\mathcal{A}$ of possible states and a different channel $\channel_a$, for each state $a\in\mathcal{A}$. The communication is governed by a sequence of states $\bm{a}=(a_1,\ldots,a_n)\in\mathcal{A}^n$ (known to both encoder and decoder)  with composition $P_n$, which determines the channels to use. In particular, channel $\channel_{a_i}$ is used at time instant $i$. The composition constraint in this case is that all codewords have conditional composition $V_n$ given $\bm{a}$, which means that any codeword has a symbol $x$ in a fraction $V_n(x|a)$ of the $nP_n(a)$ positions where $a_i=a$. We then assume that, as $n\to\infty$, $P_n\to P$ and $V_n\to V$.
\begin{remark}
Note that this general scenario includes the ordinary constant composition situation considered before, which is obtained for example when $P(a)=1$ for some $a$ and $\bm{a}=(a,a,\ldots,a)$. Note that it also includes the study of the parallel use of $K>1$ channels, which can be recovered by setting $P(a)=1/K, \forall a,$ and normalizing the block lengths by a factor $K$.
\end{remark}

For a given $P$ and $V$, let now $E(\{\channel_a\},R,V|P)$ be the optimal asymptotic error exponent achievable by codes with asymptotic conditional composition $V$ with respect to a sequence with asymptotic composition $P$ using the set of channels $\{\channel_a\}$, $a\in\mathcal{A}$. 
Then we have the following result.
\begin{theorem}
\label{th:cond-sphere-packing}
We have the inequality
\begin{equation}
E(\{\channel_a\},R,V|P)\leq \Espcc(\{\channel_a\},R-\varepsilon, V|P),
\end{equation}
where $\Espcc(\{\channel_a\},R,V|P)$ is defined by
\begin{align}
\Espcc(\{\channel_a\},R,V|P) & =  \sup_{\rho \geq 0} \left[ E_0^{\text{cc}}(\{\channel_a\},\rho,V|P) - \rho R\right],\\
E_0^{\text{cc}}(\{\channel_a\},\rho,V|P) & = \sum_a P(a) E_0^{\text{cc}}(\channel_a,\rho,V(\cdot|a)),
\end{align}
and $E_0^{\text{cc}}(\channel_a,\rho,V(\cdot|a))$ is the coefficient $E_0^{\text{cc}}$ of the sphere packing bound for channel $\channel_a$ with composition $V(\cdot|a)$, as defined in \eqref{eq:E0ccrhoP}.
\end{theorem}

\begin{IEEEproof}
See Appendix \ref{app:proof_condsp}.
\end{IEEEproof}

We observe that the function $\Espcc(\{\channel_a\},R,V|P)$ is finite for all rates $R> R_\infty(\{{\channel}_a\},V|P)$ where 
\begin{align}
R_\infty(\{{\channel}_a\},V|P)  & =  \lim_{\rho\to\infty} \frac{E_0^{\text{cc}}(\{\channel_a\},\rho,V|P)}{\rho}\\
& =\lim_{\rho\to\infty}  \sum_a P(a)  \frac{E_0^{\text{cc}}(\channel_a,\rho,V(\cdot|a))}{\rho}\\
& = \sum_a P(a) R_\infty({\channel}_a,V(\cdot|a)).
\end{align}
Furthermore, it is not difficult to show, using the same procedure used in Theorem \ref{th:Erinfty}, that for pure-state channels we have the inequality 
\begin{equation}
\Espcc(\{\channel_a\},R_\infty(\{{\channel}_a\},V|P),V|P)\leq R_\infty(\{{\channel}_a\},V|P).
\label{eq:CondEspRinfty}
\end{equation}

\subsection{Improvement of the Sphere-Packed Umbrella Bound}
\label{sec:impr_Espu}
We can now combine the bound derived above with the ideas presented in \cite{dalai-ISIT-2013b}, \cite{dalai-TIT-2013} and \cite{blahut-1977}, much in the same way as done in \cite{dalai-ISIT-2014a} \cite{dalai-TIT-2015}, to obtain a bound on the reliability of a channel $\channel$ using auxiliary classical-quantum channels $\{\tilde{\channel}_a\}$.  We limit here the discussion to the case of a pure-state channel with states $S_x=\ket{\psi_x}\bra{\psi_x}$ and pure-states auxiliary channels  $\{\tilde{\channel}_a\}$. The general case will become clear in the next section where we reformulate this bound in terms of code \emph{distances}, reinterpreting it as a generalization of the Elias bound.

For a $\rho\geq 1$, we define the set $\Gamma(\rho)$ of admissible pure-state auxiliary channels $\tilde{\channel}$ with states $\tilde{S}_x=\ket{\tilde{\psi}_{x}}\bra{\tilde{\psi}_{x}}$ such that
\begin{equation}
|\braket{\tilde{\psi}_{x}}{\tilde{\psi}_{x'}}|\leq |\braket{\psi_{x}}{\psi_{x'}}|^{1/\rho} , \quad \forall x,x' \in \mathcal{X}.
\end{equation}
For any $a\in\mathcal{A}$ we choose an auxiliary pure state channel $\tilde{\channel}_a\in\Gamma(\rho)$ with states $\tilde{S}_{a,x}=\ket{\tilde{\psi}_{a,x}}\bra{\tilde{\psi}_{a,x}}$. Given a sequence $\bm{a}=(a_1,\ldots,a_n)\in\mathcal{A}^n$ and a sequence $\bm{x}=(x_1\ldots,x_n)\in\mathcal{X}^n$, let 
\begin{equation}
\tilde{\bm{\psi}}_{\bm{a},\bm{x}} = \tilde{\psi}_{a_1,x_1}\otimes\cdots\otimes \tilde{\psi}_{a_n,x_n}.
\end{equation}
Now, given two sequences $\bm{x}=(x_1,\ldots,x_n)$ and $\bm{x}'=(x'_1,\ldots,x'_n)$, we can use these auxiliary channels to bound the overlap $ |\braket{\bm{\psi}_{\bm{x}}}{\bm{\psi}_{\bm{x}'}}|^2$  as
\begin{equation}
 |\braket{\bm{\psi}_{\bm{x}}}{\bm{\psi}_{\bm{x}'}}|^2 \geq |\braket{\tilde{\bm{\psi}}_{\bm{a},\bm{x}}}{\tilde{\bm{\psi}}_{\bm{a},\bm{x}'}}|^{2\rho}.
\end{equation}
This will allow us to bound $E(R,P)$ for the original channel using the bound (see for example \cite[Th. 12]{dalai-TIT-2013})
\begin{align}
E(R,P) & \leq -\frac{1}{n}\log \max_{m\neq m'} |\braket{\bm{\psi}_{\bm{x}_m}}{\bm{\psi}_{\bm{x}_{m'}}}|^2 + o(1)
\label{eq:ineqEinprod}\\
& \leq -\frac{\rho}{n}\log \max_{m\neq m'}|\braket{\tilde{\bm{\psi}}_{\bm{a},\bm{x}_m}}{\tilde{\bm{\psi}}_{\bm{a},\bm{x}_{m'}}}|^2 + o(1).
\label{eq:ineq1}
\end{align}
We could use the extension of the sphere packing bound considered in this section to upper bound the right hand side of the last equation as done in \cite[Sec. VIII]{dalai-TIT-2013} if all codewords $\bm{x}_m$ had the same conditional composition given the sequence $\bm{a}$. Since the sequence $\bm{a}$ is arbitrary, we choose it so that this condition is met by at least a large enough subset $\mathcal{T}$ of codewords, and we only apply the sphere packing bound to this subset $\mathcal{T}$. In order to do this, we adopt an idea proposed by Blahut \cite{blahut-1977} in a generalization of the Elias bound and already considered for a further generalization in \cite{dalai-ISIT-2014a, dalai-TIT-2015}.

Given a code with $M=e^{nR_n}$ codewords of composition $P_n$, assume that there exists a conditional composition $\hat{V}_n(a|x):\mathcal{X} \mapsto \mathcal{A}$ (i.e., $nP_n(x)\hat{V}_n(a|x)$ is an integer) such that
\begin{equation}
R_n> I(P_n,\hat{V}_n),
\end{equation}
where $I(P_n,\hat{V}_n)$ is the mutual information with the notation of \cite{csiszar-korner-book}. 
Define then 
\begin{equation}
\hat{P}_n(a)=\sum_{x}P_n(x)\hat{V}_n(a|x)
\end{equation}
(that we will write as $P_n\hat{V}_n=\hat{P}_n$) and
and let $V_n(x|a)=P_n(x)\hat{V}_n(a|x)/\hat{P}_n(a)$, so that $\hat{P}_nV_n=P_n$. Note that $I(P_n,\hat{V}_n)=I(\hat{P}_n,V_n)$.

Then, (see \cite[proof of Th. 8]{blahut-1977}, or \cite[Lemma 3]{dalai-TIT-2015}) there is at least one sequence $\bm{a}$ of composition $\hat{P}_n$ such that there is a subset $\mathcal{T}$ of at least $|\mathcal{T}|=e^{n(R_n- I(\hat{P}_n,V_n)-o(1))}$ codewords with conditional composition $V_n$ given $\bm{a}$. 
Since we are interested in the limit as $n\to\infty$, we directly work with the asymptotic rate $R$, compositions $P$ and $\hat{P}$ and matrix $V$, and we neglect the constraint that $nP_n(x)$, $nP_n(x)\hat{V}_n(a|x)$ etc. are integers.

Now, we can use the conditional sphere packing bound introduced in this section to bound the probability of error of the subcode $\mathcal{T}$ of rate $\tilde{R}=R- I(\hat{P}_n,V_n)-o(1)$ used over the varying channel $\tilde{\channel}_{a_1},\cdots,\tilde{\channel}_{a_n}$.  For these codewords used over this varying channel, there is a decision rule such that (\cite{holevo-2000}, \cite[Sec. VIII]{dalai-TIT-2013})
\begin{align}
\tilde{\mathsf{P}}_{\text{e,max}} & \leq (|\mathcal{T}|-1)\max_{m\neq m'\in \mathcal{T}}|\braket{\tilde{\bm{\psi}}_{\bm{a},\bm{x}_m}}{\tilde{\bm{\psi}}_{\bm{a},\bm{x}_{m'}}}|^2\\
& \leq e^{n(R-I(\hat{P},V)+o(1))}\max_{m\neq m'\in \mathcal{T}}|\braket{\tilde{\bm{\psi}}_{\bm{a},\bm{x}_m}}{\tilde{\bm{\psi}}_{\bm{a},\bm{x}_{m'}}}|^2.
\label{eq:ineq2}
\end{align}
On the other hand, as $n\to\infty$, Theorem \ref{th:cond-sphere-packing} with rate $\tilde{R}$ gives
\begin{align}
-\frac{1}{n}\log \tilde{\mathsf{P}}_{\text{e,max}}  & \leq \Espcc(\{\tilde{\channel}_a\},\tilde{R}-\varepsilon, V|\hat{P}) + o(1) \\
& \leq \Espcc(\{\tilde{\channel}_a\},R-I(\hat{P},V)-\varepsilon, V|\hat{P})+ o(1).
\label{eq:ineq3}
\end{align}
Putting together equations \eqref{eq:ineq1}, \eqref{eq:ineq2} and \eqref{eq:ineq3}, we obtain
\begin{equation}
E(R,P)\leq \rho [\Espcc(\{\tilde{\channel}_a\},R-I(\hat{P},V)-\varepsilon, V|\hat{P}) +R-I(\hat{P},V)].
\end{equation}
Since the choice of $\rho$, of the channels $\{\tilde{\channel}_a\}\in\Gamma(\rho)$ and of the distributions $\hat{P},V$ can be optimized, we have, in analogy with \cite[Th. 11]{dalai-TIT-2013},
\begin{theorem}
For a pure-state channel, the reliability function with constant composition $P$ satisfies $E(R,P)\leq E_{\text{spu}}^{\text{cc}}(R,P)$ where
\begin{equation}
E_{\text{spu}}^{\text{cc}}(R,P)=\inf \rho [\Espcc(\{\tilde{\channel}_a\},R-I(\hat{P},V)-\varepsilon, V|\hat{P}) +R-I(\hat{P},V)],
\label{def:Espucc}
\end{equation}
the infimum being over $\varepsilon>0$, $\rho\geq 1$, auxiliary pure-state channels $\tilde{\channel}_a\in\Gamma(\rho)$, and auxiliary distributions $\hat{P}$ and $V$ such that $\hat{P}V=P$.
\label{Th:Espucc}
\end{theorem}
\begin{remark}
Note that for the choice $\mathcal{A}=\mathcal{X}$, $V(a|x)=P(a)$, $\forall a$, we have $I(P,V)=0$. We can also notice that the optimization of the channels $\tilde{\channel}_a$ will give $\tilde{\channel}_a=\tilde{\channel}$, $\forall a$, for an optimal $\tilde{\channel}$. With this constraint on $V$, the bound $E(R,P)$ is weakened to
\begin{equation}
\inf \rho [\Espcc(\tilde{\channel},R-\varepsilon, P) +R],
\end{equation}
where the infimum is now only over $\rho\geq 1$ and $\tilde{\channel}\in \Gamma(\rho)$. This is a constant composition version of the bound in \cite[Th. 11]{dalai-TIT-2013}.
\end{remark}

\subsection{Connection with the Elias Bound}
In the same way as \cite[Th. 11]{dalai-TIT-2013} generalizes the results of \cite[Sec. III]{dalai-TIT-2013}, it possible to reinterpret the idea used to obtain Theorem \ref{Th:Espucc} as a generalization of the Elias bound presented in \cite{dalai-ISIT-2014a} and \cite{dalai-TIT-2015}.
For this purpose, it is useful to introduce a notion of distance between symbols and distance between sequences, and then restate our bound as a bound on the minimum distance of codes. Finally, bounds on the reliability function can be obtained by relating the minimum distance to the probability of error (see \cite[Sec. VI]{dalai-TIT-2015} for details).

Let $d$ be a function $d:\mathcal{X}\times\mathcal{X}\to\mathbb{R}^+\cup \{\infty\}$ such that 
\begin{align*}
d(x,x') & \geq 0\\
d(x,x')&=d(x',x) \quad \forall x,x'\\
d(x,x) & = 0.
\end{align*} 
We call this function $d$ a ``distance'' although, as seen above, we do not really require all the properties of a distance. We stress that $d$ is allowed to take value $\infty$ for some pairs of symbols, a case which is of practical interest in our context.
We extend the distance to sequences of symbols defining, for $\bm{x}=(x_1,\ldots,x_n)$ and $\bm{x}'=(x_1',\ldots,x_n')$,
\begin{equation}
d(\bm{x},\bm{x}'):=\sum_{i=1}^n d(x_i,x_i').
\end{equation}
Note in particular that $d(\bm{x},\bm{x}')=\infty$ iff $d(x_i,x_i')=\infty$ for at least one $i$.
 
For a given code $\Code$, we define its minimum distance as
\begin{equation}
\dmin(\Code) := \min_{\bm{x}, \bm{x}'\in\Code,\,\bm{x}\neq \bm{x}'} d(\bm{x},\bm{x}').
\end{equation}
For a composition $P$, we define 
\begin{equation}
d(R,n,P):=\max_{\Code} \dmin(\Code),
\end{equation}
where the maximum is over all codes of length $n$, rate at least $R$, and composition $P$. For a fixed $R$, we then define 
\begin{equation}
\delta^*(R,P):=\limsup_{n\to\infty, \{P_n\}} \frac{1}{n} d(R_n,n, P_n),
\end{equation}
where $R_n\to R$ and $P_n\to P$ as $n\to\infty$.

Note that we can drop the constant composition constraint defining
\begin{equation}
d(R,n):=\max_{\Code} \dmin(\Code),
\end{equation}
and, correspondingly,
\begin{equation}
\delta^*(R):=\limsup_{n\to\infty} \frac{1}{n} d(R,n).
\end{equation}
Then we have
\begin{equation}
\delta^*(R):=\max_P  \delta^*(R,P).
\end{equation}

We want to use our results to bound the quantity $\delta^*(R,P)$. In order to do this we proceed in a similar way as done in Section \ref{sec:impr_Espu}. Note that this corresponds to what done in  \cite{dalai-TIT-2015} with two variations; 1) we use general auxiliary classical-quantum channels in place of the so called representations composed of vectors, and 2) we replace the Lov\'asz-like trick of \cite[Lemma 2]{dalai-TIT-2015} with the sphere packing bound.

Given the distance $d$ and a $\rho\geq 1$, we define the set $\Gamma(\rho)$ of admissible auxiliary channels $\tilde{\channel}$ with states $\tilde{S}_x$ such that
\begin{equation}
\Tr \sqrt{\tilde{S}_x}\sqrt{\tilde{S}_{x'}} \leq e^{-d(x,x')/\rho}.
\end{equation}
We then consider again as in Section \ref{sec:impr_Espu} the subcode $\mathcal{T}$ of codewords with composition $P_n$ all with the same conditional composition $V_n$ given the sequence $\bm{a}$.
For any $a\in\mathcal{A}$ we choose an auxiliary channel $\tilde{\channel}_a\in\Gamma(\rho)$ with states $\tilde{S}_{a,x}$ and for an $\bm{x}\in\mathcal{T}$ we define
\begin{equation}
\tilde{\bm{S}}_{\bm{a},\bm{x}} = \tilde{S}_{a_1,x_1}\otimes\cdots\otimes \tilde{S}_{a_n,x_n}.
\end{equation}
Note that this implies that for two sequences $\bm{x}$ and $\bm{x}'$,
\begin{equation}
\Tr\sqrt{\tilde{\bm{S}}_{\bm{a},\bm{x}}}\sqrt{\tilde{\bm{S}}_{\bm{a},\bm{x'}}}\leq e^{-d(\bm{x},\bm{x'})/\rho}.
\label{eq:elias_comp_1}
\end{equation}
Consider now an optimal decision scheme for the states associated to the subcode $\mathcal{T}$, that is, $\tilde{\bm{S}}_{\bm{a},\bm{x}}$, $\bm{x}\in\mathcal{T}$. The extension of \eqref{eq:ineq2} \cite{holevo-2000} says that for such a set of states, there exists a measurement such that
\begin{equation}
\tilde{\mathsf{P}}_{\text{e,max}}  \leq e^{n(R-I(\hat{P},V)+o(1))}\max_{m\neq m'\in \mathcal{T}} \Tr\sqrt{\tilde{\bm{S}}_{\bm{a},\bm{x}_m}}\sqrt{\tilde{\bm{S}}_{\bm{a},\bm{x}_{m'}}}.
\label{eq:elias_comp_2}
\end{equation}
But, again, we can use the conditional sphere packing bound to lower bound the probability of error of the subcode $\mathcal{T}$ as
\begin{equation}
-\frac{1}{n}\log \tilde{\mathsf{P}}_{\text{e,max}}  \leq \Espcc(\{\tilde{\channel}_a\},R-I(\hat{P},V)-\varepsilon, V|\hat{P})+ o(1).
\label{eq:elias_comp_3}
\end{equation}
Combining equations \eqref{eq:elias_comp_1}, \eqref{eq:elias_comp_2} and \eqref{eq:elias_comp_3} we obtain
\begin{equation}
\frac{1}{n}\min_{m\neq m'}d(\bm{x}_m, \bm{x}_{m'}) \leq \rho(\Espcc(\{\tilde{\channel}_a\},R-I(\hat{P},V)-\varepsilon, V|\hat{P}) +R-I(\hat{P},V)) +o(1),
\end{equation}
which asymptotically gives the following result.
\begin{theorem}
\label{th:dist_elias_Espu}
For a distance $d$ and assuming the above definitions, we have the inequality
\begin{equation}
\delta^*(R,P)\leq E_{\text{spu}}^{\text{cc}}(R,P),
\end{equation}
where $E_{\text{spu}}^{\text{cc}}(R,P)$ is defined in \eqref{def:Espucc}.
\end{theorem}

As mentioned, this bound is an extension of \cite[Th. 6]{dalai-TIT-2015}.
To see this, we can consider the particular case in which we restrict the attention to pure-state auxiliary channels with states $\tilde{S}_{a,x}=\ket{\tilde{\psi}_{a,x}}\bra{\tilde{\psi}_{a,x}}$ and then study the smallest rate for which the bound $E_{\text{spu}}^{\text{cc}}(R,P)$ (with this additional constraint) is finite. First note that for fixed channels $\{\tilde{\channel}_a\}$, distributions $\hat{P}$ and $V$, and $\varepsilon$ sufficiently small, the quantity on the right hand side of equation \eqref{def:Espucc} is finite for $R>R_\infty(\{\tilde{\channel}_a\},V|\hat{P})+I(\hat{P},V)$. Furthermore, when $R$ approaches this value from the right,  using equation \eqref{eq:CondEspRinfty}, 
the right hand side of equation \eqref{def:Espucc} is upper bounded by $2\rho R_\infty(\{\tilde{\channel}_a\},V|\hat{P})$. 
So, for $R>R_\infty(\{\tilde{\channel}_a\},V|\hat{P})+I(\hat{P},V)$ we have the bound
\begin{equation}
\delta^*(R,P)  \leq 2 \rho R_\infty(\{\tilde{\channel}_a\},V|\hat{P}).
\label{eq:dbound_rinfty}
\end{equation}
For pure state auxiliary channels we can write
\begin{align}
R_\infty(\{{\channel}_a\},V|\hat{P})  & =   \sum_a P(a) R_\infty({\channel}_a,V(\cdot|a)) \\
& =   \sum_{a\in \mathcal{X}} \hat{P}(a) \min_{F_a} \left[-\sum_x V(x|a) \log\Tr(\tilde{S}_{a,x}^0 F_a)\right]\\
& =  \min_{\{F_a\}} \sum_{a,x\in \mathcal{X}} \hat{P}(a)  V(x|a) \log \frac{1}{\bra{\tilde{\psi}_{a,x}} F_a \ket{\tilde{\psi}_{a,x}}}\\
 & \leq \min_{\{f_a\}} \sum_{a,x\in \mathcal{X}} \hat{P}(a)  V(x|a) \log \frac{1}{|\braket{\tilde{\psi}_{a,x}} {f_a}|^2},
\end{align}
where the last step we have enforced minimization over rank one operators $F_a=\ket{f_a}\bra{f_a}$.
Optimizing now over $\rho$, $\hat{P}$ and $V$ such that $\hat{P}V=P$, and the auxiliary vectors $\{\tilde{\psi}_{a,x}\}$, and comparing with the definition of $\vartheta(\rho,V|\hat{P})$ used in \cite{dalai-TIT-2015}, we deduce that the bound of Theorem \ref{th:dist_elias_Espu} includes, as a particular case, the bound presented in \cite[Th. 6]{dalai-TIT-2015} as a generalization of the Elias bound for general, possibly infinite, distances\footnote{Note that the definition of $\Gamma(\rho)$  in \cite{dalai-TIT-2015} is slightly different than here, so that the parameter $\rho$ here corresponds to the parameter $\rho/2$ there.}. Hence, it includes in particular all previously known extensions as discussed in \cite{dalai-TIT-2015}.

\section{Acknowledgments}
The author(s) would like to thank the Isaac Newton Institute for Mathematical Sciences, Cambridge, for support and hospitality during the programme ``Mathematical Challenges in Quantum Information'' where work on this paper was undertaken. AW was supported by the European Commission (STREP ``RAQUEL''),  the ERC (Advanced Grant ``IRQUAT''),​ the Spanish MINECO (grant FIS2013-40627-P) with the​ support of FEDER funds, and by the Generalitat de​ Catalunya CIRIT, project 2014-SGR-966.

\appendices
\section{Proof of Theorem \ref{th:cc-sphere-packing}}
\label{app:proof_sp}

The structure of the proof is the same as in \cite{shannon-gallager-berlekamp-1967-1}, and \cite[Th. 5]{dalai-TIT-2013} with some technical changes which are required for dealing with general compositions. While introducing this changes, we also considerably simplify some of the technicalities with respect to \cite[Th. 5]{dalai-TIT-2013} in order to give a simpler yet more transparent proof of both this and the original theorem. 

From the definition of $E(R,P)$, there exists a sequence of codes of block-lengths $n=1,2,\ldots$ with rates $R_n\to R$, compositions $P_n\to P$ and with probabilities of error $\Pemax^{(n)}$ such that
\begin{equation*}
E(R,P)=\limsup_{n\to\infty} -\frac{1}{n}\log \Pemax^{(n)}.
\end{equation*}
We first observe that we can just focus on the subset of input symbols with $P(x)>0$ and assume without loss of generality that $P_n(x)=0$ if $P(x)=0$. This technicality is needed after equation \eqref{eq:defFs} below and can be motivated as follows. Let $\mathcal{X}_0$ be the subset of $\mathcal{X}$ such that $P(x)=0$ if and only if $x\in\mathcal{X}_0$. Then, for for any sequence of compositions $P_n\to P$, for all $x\in\mathcal{X}_0$ we can write that $P_n(x)\leq \varepsilon_n/|\mathcal{X}_0|$, where $\varepsilon_n\to 0$ as $n\to \infty$. Any codeword with composition $P_n$ will contain symbols in $\mathcal{X}_0$ in at most $n\varepsilon_n $ positions. There are only nearly $e^{n H(\varepsilon_n )}$ choices for these positions and, for each such choice there are only at most $|\mathcal{X}_0|^{n\varepsilon_n }$ possible combinations of symbols in $\mathcal{X}_0$. Hence, from a code with rate $R_n$ and composition $P_n$ we can extract a subcode with rate $R_n'=R_n-H(\varepsilon_n)-\varepsilon_n \log |\mathcal{X}_0|$ such that each symbol in $\mathcal{X}_0$ appears precisely in the same positions in all codewords. We can then bound $E(R,P)$ by bounding the probability of error for this subcode since, given that $\varepsilon_n\to 0$, we have $(R_n'-R_n) \to 0$.
However, in the chosen subcode each symbol in $\mathcal{X}_0$ appears in the same positions in all codewords, and can thus be replaced with any symbol in $\mathcal{X}\backslash \mathcal{X}_0$ without affecting the probability of error.

For every fixed $n$, the idea is again as in previous proofs to consider a binary hypothesis test between a properly selected code signal $\bm{S}_{\bm{x}_m}$ and an auxiliary density operator $\bm{F}=F^{\otimes n}$. The main difference with respect to \cite[Th. 5]{dalai-TIT-2013} is in the choice of $F$ and, as a consequence, in some technical details. 

Let $n$ be fixed and let $M$ be the number of codewords, that is $M=e^{n R_n}$. For any $m=1,\ldots,M$ consider a binary hypothesis test between $\bm{S}_{\bm{{x}}_m}$ and an auxiliary state $\bm{F}=F^{\otimes n}$. We assume that the supports of the two operators are not disjoint and, with the notation used in \cite{dalai-TIT-2013}, we define the quantity
\begin{align}
\mu(s) & =   \mu_{\bm{S}_{\bm{{x}}_m},\bm{F}}(s)\nonumber\\
& = \log \Tr \bm{S}_{\bm{{x}}_m}^{1-s}\bm{F}^s\nonumber.
\end{align}
Note that, setting
\begin{equation}
\mu_{S_x,F}(s)=\log\left(\Tr S_{{x}}^{1-s} F^s\right),
\end{equation}
we can write 
\begin{align}
\mu_{\bm{S}_{\bm{{x}}_m},\bm{F}}(s) & =  \log \prod_{i=1}^n \Tr S_{{x}_{m,i}}^{1-s} F^s \nonumber\\
 & =  \log \prod_{{x}} \left(\Tr S_{{x}}^{1-s} F^s\right)^{n P_n({x})}\nonumber\\
 & =  n \sum_{x} P_n({x})\mu_{S_x,F}(s).
 \label{eq:muconstcomp}
\end{align}
Applying \cite[Th. 4]{dalai-TIT-2013}, we find that for each $s$ in $(0,1)$, either 
\begin{equation}
\Tr\left[\left(\mathds{1}-\Pi_m\right)\bm{S}_{\bm{{x}}_m}\right]>\frac{1}{8}\exp\left[\mu(s)-s\mu'(s)-s\sqrt{2\mu''(s)}\right]
\label{eq:BinHypSP-1}
\end{equation}
or
\begin{equation}
\Tr\left[ \Pi_m\bm{F}\right]>\frac{1}{8}\exp\left[\mu(s)+(1-s)\mu'(s)-(1-s)\sqrt{2\mu''(s)}\right].
\label{eq:BinHypSP-2}
\end{equation}
As in \cite[Th. 5]{dalai-TIT-2013}, this can be converted in a relation between $\Pemax^{(n)}$ and $R_n$ in the form that either
\begin{equation}
\Pemax^{(n)}>\frac{1}{8}\exp\left[\mu(s)-s\mu'(s)-s\sqrt{2\mu''(s)}\right]
\label{eq:cond1}
\end{equation}
or
\begin{equation}
R_n<-\frac{1}{n}\left[ \mu(s)+(1-s)\mu'(s)-(1-s)\sqrt{2\mu''(s)} - \log 8\right].
\label{eq:cond2}
\end{equation}

Note that due to \eqref{eq:muconstcomp}, the right hand side of \eqref{eq:cond2} only depends on $n$, $s$, $P_n$, and $F$. Let then this quantity be called $R_n(s,P_n,F)$, that is,
\begin{equation}
R_n\left(s,P_n,F\right)=-\frac{1}{n}\Bigl( \mu(s)+(1-s)\mu'(s) \Bigr. \Bigl. - (1-s)\sqrt{2\mu''(s)} - \log 8 \Bigr). 
\end{equation}
We can use this equation to write $\mu'(s)$ in terms of $R_n(s,P_n,F)$. Using \eqref{eq:muconstcomp}, we can state our conditions by saying that either
\begin{equation}
R_n<R_n(s,P_n,F)
\label{eq:condRnRnsPnF}
\end{equation}
or
\begin{equation}
\frac{1}{n}\log \frac{1}{\Pemax^{(n)}} < -\frac{1}{1-s}\sum_{x} P_n({x}) \mu_{S_{x},F}(s) -\frac{s}{1-s} R_n(s,P_n,F)
+\frac{1}{n}\left(2s\sqrt{2\mu''(s)} +\frac{\log 8}{1-s}\right).
\label{eq:cond1/nPemax<..sumqkmuk}
\end{equation}

At this point we introduce the variation with respect to \cite{dalai-TIT-2013}. For any $F$, one of the two conditions above must be satisfied and, in \cite{dalai-TIT-2013}, the choice of $F$ was made which guaranteed the best bound for the \emph{optimal} compositions $P_n$. Here, instead, the compositions $P_n$ are forced to tend to a given composition $P$ and we have to choose $F$ accordingly. For a given $s\in (0,1)$, let $F_s$ be the operator defined by
\begin{equation}
F_s = \argmin_F - \sum_x P(x) \log(\Tr S_x^{1-s}F^s).
\label{eq:defFs}
\end{equation}
Note that this choice guarantees that for all $x$ with $P(x)>0$, $S_x$ and $F$ have non-disjoint supports. Since we assumed that $P_n(x)=0$ whenever $P(x)=0$, the requirement that
$\bm{S}_{\bm{{x}}_m}$ and $\bm{F}$ have non-disjoint support is satisfied for all sequences ${\bm{{x}}_m}$ with composition $P_n$, and hence $\mu(s)$ is a finite quantity for all $s\in(0,1)$.

 We will now relate the choice of $s$ to the rate $R$ and then use $F_s$ in place of $F$ for the chosen $s$ (it must be clear, however, that $\mu'(s)$ and $\mu''(s)$ are computed by holding $F$ fixed). Note that we can write
\begin{equation}
R_n(s,P_n,F_s)=- \sum_{x} P_n({x})\left[ \mu_{S_{x},F_s}(s)+(1-s) \mu_{S_{x},F_s}'(s)\right]
+\frac{1}{\sqrt{n}}(1-s)\sqrt{2\sum_{x} P_n({x}) \mu_{S_{x},F_s}''(s)} + \frac{1}{n} \log 8.
\label{eq:defRnfinal}
\end{equation}
For any fixed $s$, the last two terms on the right hand side vanish as $n\to\infty$, and $P_n$ in the first term tends to $P$. Hence, it is useful to define the quantity
\begin{align}
R^*(s,P) & = \lim_{n\to\infty}R_n(s,P_n,F_s)\\
& = - \sum_{x} P({x})\left[ \mu_{S_{x},F_s}(s)+(1-s) \mu_{S_{x},F_s}'(s)\right]
\end{align}
and compare this quantity to the rate $R$ which we are considering, which is the limit of the $R_n$'s.

We first observe that, for any $x$ and $F$, $\mu_{S_{x},F}(s)$ is a non-positive convex function of $s$ for all $s\in (0,1)$, which implies that for any $F$ we have 
\begin{align*}
\mu_{S_{x},F}(s)+ (1-s) \mu_{S_{x},F}'(s) & \leq \mu_{S_{x},F}(1^-)\\ & \leq 0.
\end{align*}
Hence, both $R^*(s,P)$ and $R_n(s,P_n,F_s)$ are non-negative quantities. Furthermore, it is not difficult to see that $F_s$ is continuous in $s$ in the interval $0<s<1$, and so is $R^*(s,P)$. 
Hence, $R^*(s,P)$ is a continuous non-negative function of $s$ in the interval $0<s<1$, and we can compare this function with the asymptotic rate $R$.
We only have three possible situations:
\begin{enumerate}
\item $R>\sup_{s\in (0,1)} R^*(s,P)$;
\item $R\leq \inf_{s\in(0,1)}R^*(s,P)$;
\item $\inf_{s\in(0,1)}R^*(s,P)<R\leq \sup_{s\in(0,1)}R^*(s,P)$.
\end{enumerate}

Assume case 1) is verified. Fix an arbitrary $s\in(0,1)$. Since $R_n\to R$ and $R_n(s,P_n,F_s)\to R^*(s,P)<R$, $R_n>R_n(s,P_n,F_s)$ for all $n$ large enough. Hence, equation \eqref{eq:condRnRnsPnF} is not satisfied and thus equation \eqref{eq:cond1/nPemax<..sumqkmuk} is. 
Since $s$ is fixed and $R_n(s,P_n,F_s)\geq 0$, as $n$ goes to infinity we find
\begin{align}
\frac{1}{n}\log \frac{1}{\Pemax^{(n)}} & <  -\frac{1}{1-s}\sum_{x} P_n({x}) \mu_{S_{x},F_s}(s) -\frac{s}{1-s} R_n(s,P_n,F_s)+o(1)\\
& \leq -\frac{1}{1-s}\sum_{x} P_n({x}) \mu_{S_{x},F_s}(s)+ o(1)
\end{align}
and in the limit, since $P_n\to P$,
\begin{align}
E(R,P) 
& \leq E_0^{\text{cc}}\left(\frac{s}{1-s},P\right).
\label{eq:E(R)to0}
\end{align}
Since this holds for arbitrary $s\in (0,1)$, we have
\begin{align*}
E(R,P) & \leq \lim_{s \to 0} E_0^{\text{cc}}\left(\frac{s}{1-s},P\right) \\
& = 0,
\end{align*}
where the last step is deduced by noticing that $E_0^{\text{cc}}(\rho,P)$ is continuous at $\rho=0$ and that the argument of the minimization in the definition of $E_0^{\text{cc}}(\rho,P)$ is a non-negative quantity which, for $\rho=0$, vanishes for all $F$ with full support\footnote{Note, however, that for $\rho>0$ there is a unique optimal $F$, which makes $F_s$ well defined.}. This proves the theorem in case 1) since $\Espcc(R-\varepsilon,P)\geq 0$.

Assume now that case 2) is satisfied, which means by definition of $R^*(s,P)$ that, for any $s\in(0,1)$, we have
\begin{align*}
R \leq  - \sum_{x} P({x})\left[ \mu_{S_{x},F_s}(s)+(1-s) \mu_{S_{x},F_s}'(s)\right].
\end{align*}
Now, since $\mu_{S_{x},F}(s)$ is convex and non-positive for all $F$, it is possible to observe that $\mu_{S_{x},F_s}(s)-s\mu_{S_{x},F_s}'(s)\leq 0$, which implies that $-\mu_{S_{x},F_s}'(s)\leq -\mu_{S_{x},F_s}(s)/s$. Thus, for all $s\in(0,1)$, 
\begin{align*}
R & \leq  \sum_x P(x) \left(- \frac{1}{s}\mu_{S_x,F_s}(s)\right)\\
& \leq  \frac{1-s}{s}E_0^{\text{cc}}\left(\frac{s}{1-s},P\right).
\end{align*}
Calling now $\rho=s/(1-s)$, we find that for all $\rho>0$
\begin{equation*}
R  \leq  \frac{E_0^{\text{cc}}(\rho,P)}{\rho}.
\end{equation*}
Hence, for any $\varepsilon>0$, we find
\begin{align*}
\Espcc\left(R - \varepsilon,P\right) & =  \sup_{\rho>0}\left(E_0^{\text{cc}}\left(\rho,P\right) -\rho \left( R-\varepsilon\right) \right)
 \\   & \geq  \sup_{\rho>0} (\rho\, \varepsilon ).
\end{align*}
This means that $\Espcc(R-\varepsilon,P)$ is unbounded for any $\varepsilon>0$, which obviously implies that $E(R,P)\leq \Espcc(R-\varepsilon,P)$ for all positive $\varepsilon$, proving the theorem in this case.

Finally, assume that case 3) above is satisfied. Then, for any $\varepsilon>0$ small enough, there is a $\bar{s}$ such that $R^*(\bar{s},P)=R-\varepsilon$. For this fixed value $\bar{s}$, since again $R_n\to R$ and $R_n(\bar{s},P_n,F_{\bar{s}})\to R^*(\bar{s},P)=R-\varepsilon$, $R_n>R_n(\bar{s},P_n,F_{\bar{s}})$ for all $n$ large enough. Hence, for $s=\bar{s}$, for all $n$ large enough equation \eqref{eq:condRnRnsPnF} is not satisfied and thus \eqref{eq:cond1/nPemax<..sumqkmuk} is. This implies that, for all $n$ large enough
\begin{align}
\frac{1}{n}\log \frac{1}{\Pemax^{(n)}} & < -\frac{1}{1-\bar{s}}\sum_{x} P_n({x}) \mu_{S_{x},F_{\bar{s}}}(\bar{s}) -\frac{\bar{s}}{1-\bar{s}} R_n(\bar{s},P_n,F_{\bar{s}})
+\frac{1}{n}\left(2\bar{s}\sqrt{2\mu''(\bar{s})} +\frac{\log 8}{1-\bar{s}}\right).
\end{align} 
In the limit as $n\to\infty$ the last term vanishes, $R_n(\bar{s},P_n,F_{\bar{s}})\to R^*(\bar{s},P)=R-\varepsilon$ and $P_n\to P$. We thus conclude that 
\begin{align*}
E(R,P) & \leq -\frac{1}{1-\bar{s}}\sum_{x} P({x}) \mu_{S_{x},F_{\bar{s}}}(\bar{s}) -\frac{\bar{s}}{1-\bar{s}} (R-\varepsilon)\\
& = E_0^{\text{cc}}\left(\frac{\bar{s}}{1-\bar{s}},P\right)-\frac{\bar{s}}{1-\bar{s}} (R-\varepsilon)\\
& \leq \sup_{\rho\geq0}\left(E_0^{\text{cc}}\left(\rho,P\right) 
-\rho (R-\varepsilon)\right)\\
& =  \Espcc(R-\varepsilon,P).
\end{align*}
This holds for all $\varepsilon>0$ small enough and hence, since $\Espcc(R,P)$ is non increasing in $R$, it holds for all $\varepsilon \in (0,R)$. This concludes the proof.

\section{Proof of Theorem \ref{th:cond-sphere-packing}}
\label{app:proof_condsp}

The proof is obtained by introducing a variation in the proof of theorem \ref{th:cc-sphere-packing} presented in Appendix \ref{app:proof_sp}. In particular, we use a different operator $\bm{F}$ which we choose so as to take into account the state dependent structure of the communication process.

From the hypotheses, the communication is governed by the sequence of states $\bm{a}=(a_1,\ldots,a_n)$ with composition $P_n$, where $P_n\to P$, and codes are considered with conditional compositions $V_n$ given $\bm{a}$, where $V_n\to V$. Here again, as in the other proof, we can assume that $V_n(x|a)=0$ if $P(a)=0$ or $V(x|a)=0$.
The structure of the proof remains unchanged with the only difference that, instead of building $\bm{F}$ using $n$ identical copies of a single density operators $F$, we can use $|\mathcal{A}|$ different operators $F_a$, $a\in\mathcal{A}$ to build $\bm{F}$ as
\begin{equation}
\bm{F}=F_{a_1}\otimes F_{a_2} \otimes \cdots \otimes F_{a_n}.
\end{equation}
Then we can still use the two equations \eqref{eq:cond1} and \eqref{eq:cond2} to bound the probability of error as a function of the rate, with the difference that the function $\mu(s)$ now reads
\begin{equation}
\mu_{\bm{S}_{\bm{{x}}_m},\bm{F}}(s) =  n \sum_{a, x} P_n({a})V_n(x|a)\mu_{S_{x},F_a}(s).
\end{equation}
For a given $a\in\mathcal{A}$ and fixed $s$, we then choose
\begin{equation}
F_{a,s} = \argmin_F - \sum_x V(x|a) \log(\Tr S_x^{1-s}F^s), 
\label{eq:defcondFs}
\end{equation}
again ensuring that $\mu_{\bm{S}_{\bm{{x}}_m},\bm{F}}(s)$ is finite. The rest of the proof follows essentially identical with the obvious differences due to the use of quantities $E_0^{\text{cc}}(\channel_a,\rho,V(\cdot|a))$ in place of $E_0^{\text{cc}}(\rho,P)$ used before.

\section{A Remark on Haroutunian's Proof of the Shere Packing Bound}
\label{app:BHT-cl-cl-q}
As mentioned, a greedy extension of Haroutunian's proof of the sphere packing bound to quantum channels, as outlined in equation \eqref{eq:haroutunianQ}, gives a bound which is in general weak.
The reason why this happens in the quantum case and not in the classical one can be traced back to a fundamental difference in the solution to the quantum binary hypothesis testing problem in those two contexts. In fact, as seen from equations \eqref{eq:BinHypSP-1} and \eqref{eq:BinHypSP-2}, the key ingredient in the proof of the sphere packing bound is a binary hypothesis test to distinguish the state $\bm{S}_{\bm{x}_m}$ from the auxiliary state $\bm{F}$. Here, a fundamental difference with the classical counterpart is related to the roles of the Kullback-Leibler discrimination and Renyi divergence in the expression for the error exponents in binary hypothesis testing. This difference was already observed in \cite[Sec. 4, Remark 1]{nagaoka-2006} and \cite[Sec. 4.8]{audenaert-et-al-2008} and leads to the mentioned difference in the expressions for the sphere packing bound. We discuss it here in detail for completeness.

In a binary hypothesis testing between two density operators $A$ and $B$, based on $n$ independent extractions, the error exponents of the first and second kind can be expressed parametrically as (see \cite{audenaert-et-al-2008}, \cite{dalai-TIT-2013})
\begin{align}
-\frac{1}{n}\log {\Pe}_{|A} & =-\mu(s)+s\mu'(s)+o(1)\\
-\frac{1}{n}\log {\Pe}_{|B} & =-\mu(s)-(1-s)\mu'(s)+o(1)
\end{align}
where 
\begin{equation}
\mu(s)=\log \Tr A^{1-s}B^s.
\end{equation}
Upon differentiation, one finds
\begin{align}
-\frac{1}{n}\log {\Pe}_{|A} & =-\log\Tr(A^{1-s}B^s) + \Tr\left[\frac{A^{1-s}B^s}{\Tr A^{1-s}B^s}\left(\log B^s-\log A^s \right)\right]+o(1)\\
-\frac{1}{n}\log {\Pe}_{|B} & =-\log\Tr(A^{1-s}B^s) + \Tr\left[\frac{A^{1-s}B^s}{\Tr A^{1-s}B^s}\left(\log A^{1-s}-\log B^{1-s} \right)\right]+o(1)
\end{align}
In the classical case, $A$ and $B$ commute. We can then define the density operator $V_s=\frac{A^{1-s}B^s}{\Tr A^{1-s}B^s}$ and use the properties $\log B^s-\log A^s=\log A^{1-s}B^s -\log A $ and $\log A^{1-s}-\log B^{1-s}=\log A^{1-s}B^s-\log B$ to obtain
\begin{align}
-\frac{1}{n}\log {\Pe}_{|A} & =\Tr V_s(\log V_s-\log A)+o(1)\\
& = D(V_s||A)+o(1)
\end{align}
and 
\begin{align}
-\frac{1}{n}\log {\Pe}_{|B} & =\Tr V_s(\log V_s-\log A)+o(1)\\
 & = D(V_s||B)+o(1)
\end{align}
However, if $A$ and $B$ do not commute, the above simplification is not possible.
This discussion extends without fundamental differences to the binary hypothesis test between the state $\bm{S}_{\bm{x}_m}$ and the auxiliary state $\bm{F}$ with the exponents expressed as in equations \eqref{eq:BinHypSP-1} and \eqref{eq:BinHypSP-2}. If we assume that all the $S_x$ operators and $F$ commute, the exponents of the binary hypothesis test used in the sphere packing bound can be expressed in terms of Kullback-Leibler divergences.
For a given $s$, instead of a single density operator $V_s$ we will have a $V_{x,s}$ for each $x$, defined as $V_{x,s}=S_x^{1-s}F^s/\Tr(S_x^{1-s}F^s)$. It then turns out that the optimal $F$ to use, that is the operator $F_s$ defined in equation \eqref{eq:defFs}, is such that (see \cite[eq. (9.50)]{fano-book}, \cite[Cor. 3]{blahut-1974})
\begin{equation}
F_s=\sum_x P(x) V_{x,s}
\end{equation}
and this leads exactly to the usual expression of the sphere packing bound in terms of Kullback-Leibler expressions as in Haroutunian's roof (see in particular \cite[eq. (19)]{haroutunian-1968} and \cite[eqs. (9.23), (9.24)]{fano-book}). In the non commutative case, however, this simplification is not possible and this implies that we cannot express the sphere packing bound using the Kullback-Leibler divergence in the standard way.








%




\end{document}